\documentclass[prb,twocolumn,amsmath,amssymb]{revtex4}

\usepackage{color}
\usepackage{graphicx}
\usepackage{dcolumn}
\usepackage{bm}

\def\be{\begin{equation}}
\def\ee{\end{equation}}
\def\bd{\begin{displaymath}}
\def\ed{\end{displaymath}}
\def\-{\phantom{-}}

\begin{document}

\title{Induced polarization at a paraelectric/superconducting interface}

\author{J.T. Haraldsen}
\author{S.A. Trugman}
\author{A.V. Balatsky}
\affiliation{Theoretical Division, Los Alamos National Laboratory, Los Alamos, NM 87545, USA}
\affiliation{Center for Integrated Nanotechnologies, Los Alamos National Laboratory, Los Alamos, NM 87545, USA}

\begin{abstract}

We examine the modified electronic states at the interface between superconducting and ferro(para)-electric heterostructures. We find that electric polarization $P$ and superconducting $\psi$ order parameters can be significantly modified due to coupling through linear terms brought about by explicit symmetry breaking at the interface. Using an effective action and a Ginzburg-Landau formalism, we show that an interaction term linear in the electric polarization will modify the superconducting order parameter $\psi$ at the interface. This also produces modulation of a ferroelectric polarization. It is shown that a paraelectric-superconductor interaction will produce an interface-induced polarization.

\end{abstract}

\maketitle


\begin{figure}
\includegraphics[width=2.25in]{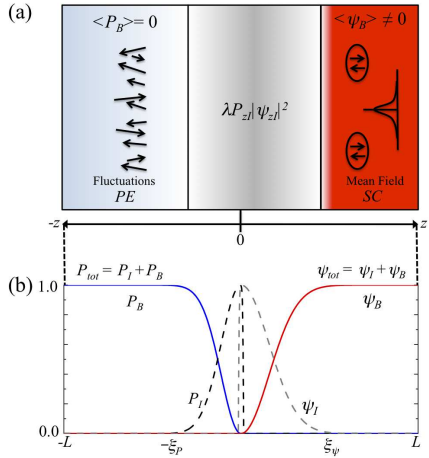}
\caption{(color online) (a) A paraelectric (PE) and superconducting (SC) interface consisting of bulk PE and SC regions, as well as an interaction region of general length $\xi_{\psi}-\xi_{P}$, where $\xi_{P}$ and $\xi_{\psi}$ are the interaction correlation lengths for the polarization and superconducting order parameters. The interaction region consists of a linear coupling of the polarization to the charge density of the superconductor. (b) The separation of the order parameters over a length $L$ and normalize to 1 to reflect the relationship between the bulk and interface regions. We use subscript $I$ and $B$ to distinguish the interaction (dashed lines) and the bulk regions (solid lines). We illustrate a slight overlap at $z$ = 0, which designates a thin film interface.}
\label{interface}
\end{figure}

{\it Introduction:} ~ In recent years, investigations into the interfaces of oxide heterostructures have revealed  that the interaction between LaAlO$_3$ and SrTiO$_3$ produces an interfacial 2-D electron gas that is superconducting at low temperatures \cite{hebe:09,ohto:02,ohto:04,reyr:07}. While this occurs at much lower temperatures than high-T$_c$ superconductors, this finding shows the importance of investigating the interfacial interactions between competing fields. While there are limitations due to lattice matching and strain within the materials,\cite{naka:06,waru:09} the coupling between these materials may provide information that is critical to the understanding of these heterostructures. 

With much of the focus of competing orders being directed at superconducting and magnetic phases including various materials from cuprates and manganites to iron pnictides,\cite{alva:05,burg:01,lums:10,chen:07,vojt:00} we look to investigate the effects of paraelectric (PE) states on superconducting (SC) order, as well as the interaction with ferroelectric (FE) order fluctuations. The phenomenological coupling of FE and SC multilayers has been extensively studied.\cite{pavl:05}  However, while it is expected that FE order will interact with the charge density of the superconductor and provide changes of the order parameters, the effects of a PE state is not well understood.\cite{pavl:05,jona:62,rabe:07}

Figure \ref{interface}(a) illustrates the PE/SC interface, where the middle area denotes the interaction correlation region of length $\xi_{\psi}-\xi_{P}$. Here, we separate the interface and bulk order parameters with subscripts $I$ and $B$ to emphasize the interaction at the interface. As shown in Fig. \ref{interface}(b), the interaction region has a specific correlation length for each order parameter, $\xi_P$ and $\xi_{\psi}$. Due to lattice imperfections, we can assume the interaction region will exhibit a Gaussian-like decay into the bulk. Within these systems, we investigate the coupling of the superconducting $\psi$ and polarization $P$ order parameters within the interface region ($\psi_I$ and $P_I$), where polarization can be an arbitrary vector. Therefore, for simplicity, we assume the polarization $P$ to be only along the $z$-axis (perpendicular to the interface). The investigation of modifications of order parameters will lead to a better understanding of the interactions and resultant phases.  

Due to the significant differences in the physics of interfaces, we can consider couplings normally not accessible in the bulk: i) explicit inversion symmetry breaking at the interface (left and right half films are different), enables an interaction that is linear in the electric polarization $P_{z_I}$ with the SC order parameter square $\Psi^2$: $S_{int} = \lambda P_{z_I}|\psi_{z_I}|^2$.\cite{linearnote} ii) Using effective action and Ginzburg-Landau formalism, we find that the interaction of a PE state and SC order produces an interface-induced polarization, while the SC state exhibits a modulation in $\psi_{z_I}$. In the approach below, we focus on electronic coupling and ignore the strain effects.

The effects of surface strain and inhomogeneities of bulk FE materials have been discussed in great detail in Ref. [\onlinecite{chan:07}]. It is shown that gradient effects can lead to the formation of multiple domains and produce a shift in $T_c$ at the surface. This produces an enhancement of the surface polarization within PE and FE materials. Our work examines the effects of a linear interface interaction in FE/SC and PE/SC heterostructures assuming homogeneous domains to gain a better perspective on these effects. In contrast, our work produces a similar enhancement of the polarization through this interface interaction. However, we also observe a modulation of the order parameters within the ordered states. Future work will examine the properties produced by interface defects and strain, inhomogeneities, and higher order coupling.

\begin{figure}
\includegraphics[width=2.75in]{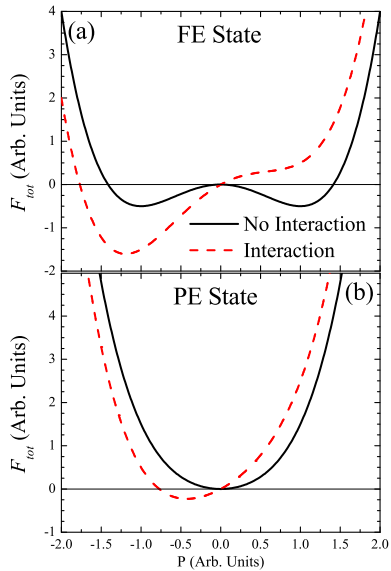}
\caption{(color online) The effect of a linear interaction term on the GL free energy for the polarization with (a) $\gamma<0$ (FE state) and (b) $\gamma>0$ (PE state). Here, the linear term provides a distinct bound polarization at the interface due to a shift of the equilibrium position of the order parameter. It is assumed that ${\gamma}$ = $\lambda$ = $\psi_0$ = 1.0, $\eta$=0.1, and $g_p$=$\rho_p$ = 0.}
\label{FE}
\end{figure}

{\it Electric Polarization and Superconductivity}:  Due to the symmetry breaking interface of heterostructures and directionality of the electric polarization, we examine the effects of a linear interaction of FE and PE states on the charge density within a superconductor. The relevant SC length scale is set by a fairly large coherence length $\xi/2$, typically on the scale of tens to hundreds of angstroms.\cite{ande:97} Therefore the use of a long wavelength approach like action and GL functional is a reasonable approach.\cite{kett:99,park:69} To investigate the interaction at the interface between an electric polarization and superconducting phases, we begin by examining the effective action given by
\be
\begin{array}{c}
S_{I} = S_{\psi_{z_I}} + S_{P_{z_I}} + S_{int}\\ \\
S_{\psi_{z_I}} = \alpha |\psi_{z_I}|^2 + \frac{\beta}{2} |\psi_{z_I}|^4 + g_{\psi}
|\bigtriangledown \psi_{z_I}|^2 +\rho_{\psi}\dot{|\psi_{z_I}|}^2\\ \\
S_{P_{z_I}} = \gamma |P_{z_I}|^2
+ \frac{\eta}{2} |P_{z_I}|^4
+ g_{p}
|\bigtriangledown P_{z_I}|^2 +\rho_{p}\dot{|P_{z_I}|}^2\\ \\
S_{int}=\lambda P_{z_I} |\psi_{z_I}|^2 \\
\end{array}
\label{G_L} \ee
where $\alpha$ = $a(T-T_{c})$ describes the SC state when $\alpha<$ 0, and $\gamma>$ 0 describes a PE state\cite{jona:62,rabe:07} and ${\bf P} = (0,0,P_z)$. The local electron density can increase or decrease depending on positive or negative $P_z$ at the interface, respectively. All other parameters are assumed to be positive constants. Here, we assume that the interface parameters are the same as the bulk. Differences in the GL parameters will also lead to other modifications of the interface. Interactions of the polarization at the interface may affect the carrier density within the superconducting state, which will either enhance or diminish $T_{c}$, which will affect the sign of $\alpha$. Due to the explicit inversion symmetry breaking at the interface, we can consider an interaction term that is linear in $P_{z_I}$, which couples the polarization to the charge density.\cite{linearnote}

Figure \ref{FE}(a) and (b) shows the effect of a linear polarization term on the GL free energy (ignoring the gradient and time-dependent terms) for the FE ($\gamma<$ 0) and PE ($\gamma>$ 0) states, respectively. Interaction terms that are quadratic in $P_{z_I}$ would simply adjust the slope of the curves. However, a linear term will shift the minima for the PE state. In the case of the FE state, the interaction will favor a specific polarization, while for the PE state, the linear term will shift the relative minima and produce a overall stabilized polarization. Since this will occur only with the interface interaction, it is expected that this effect will decay off as one moves towards the over all bulk state.

Given that we are mainly interested in investigating the effects of the PE state on the interface, we can ignore the effects of the quartic terms in the effective action since their effect is negligible to the overall state. This is because within the mean-field approximation, the quartic terms simply adds a negligible amount to the free energy with small deviations from the minima of the PE state.

\begin{figure}
\includegraphics[width=2.25in]{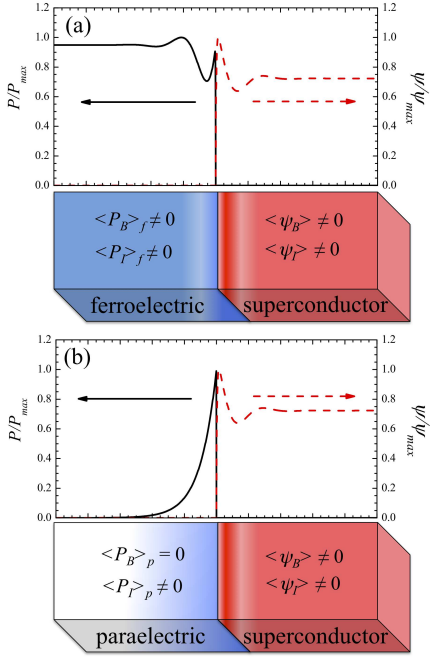}
\caption{(color online) (a) A FE (blue) and SC (red) interface with negative $\tilde{\alpha}$ and $\tilde{\gamma}$. Here, we illustrate the modulation of $\psi_z$ and $P_z$. (b) The interface of a PE (blue) and superconductor (red) with negative $\tilde{\alpha}$ and positive $\tilde{\gamma}$. The plot demonstrates the decaying interface-induced electric polarization, while the SC order parameter has a modulation at the interface. Here, it is assumed that $\tilde{\alpha}$ = $\tilde{\gamma}$ = $g$ = $\lambda$ = 1. However, the sign of $\alpha$ and $\gamma$ are as defined above. We have also multiplied the interface contributions by $e^{-|z|/\xi_d}$ to simulate dephasing caused by lattice imperfections and inhomogeneties, where $\xi_i$ for $i$ = $P$ or $\psi$ is the dephasing length for the lattice. Note: On the atomic scale, some traces of $\psi_{z}$ and $P_{z}$ may have a finite probability of exist in the opposite regions.}
\label{Pdecay}
\end{figure}

To examine the effects of the linear interaction in more detail, we examine the whole interaction region as a function of $z$ (including the gradient and time-dependent terms).  Here, the equations of motion for $P_{z_I}$ can be written as
\be
 \frac{\delta S}{\delta P_{z_I}} = (\gamma-g_{p}\bigtriangledown^2-\rho_{p}\partial_t^2)P_{z_I}+\lambda|\psi_{0}|^2 = 0, 
\ee 
or
\be 
\frac{\delta S}{\delta P_{k_I}} = (\gamma+g_{p}k_{p}^2+\rho_{p}\omega_{p}^2)P_{k_I} + \lambda |\psi_0|^2 = 0
\ee
through the use of Fourier components. The equations of motion for $\psi_{z_I}$ are similarly given by
\be
\frac{\delta S}{\delta \psi_{z_I}} = (\alpha-g_{\psi}\bigtriangledown^2-\rho_{\psi}\partial_t^2)\psi_{z_I}+2\lambda P_{0}\psi_{0} = 0
\ee
or
\be
\frac{\delta S}{\delta \psi_{k_I}} = (\alpha+g_{\psi}k_{\psi}^2+\rho_{\psi}\omega_{\psi}^2)\psi_{k_I} + 2\lambda P_0\psi_0 =0,
\ee
where $P_0$ and $\psi_0$ are the effective order parameters at the interface ($z$ = 0), which is considered as a course grain boundary integrated over the Thomas-Fermi screening length.\cite{linearnote} Here, $k$ and $
\omega$ come from the standard Fourier transforms of the spatial gradient and temporal fluctuations. From the equations of motion, it is clear that the fluctuations will add to the effective order parameters and lead to a disruption of the ordered states. We can gain a better understanding of the interface dependence of the order parameters by solving for the $P_{k_I}$ and $\psi_{k_I}$ and Fourier transform into real space. Through this, $P_{z}$ = $P_{z_I}$ + $P_{z_B}$  for $z<$0 is found to be
\be
\begin{array}{c}
\displaystyle P_{z} =  \int_{-\infty}^{\infty} \frac{-\lambda \psi_0^2}{\tilde{\gamma}+ g_pk_p^2}e^{ik_pz} dk_p + P_{z_B} \\ \\
=
\begin{cases}
\frac{\lambda \psi_0^2}{|\tilde{\gamma}| \tau_p }e^{-\frac{|z|}{\xi_p}} & \tilde{\gamma} >0 \\
\frac{\lambda \psi_0^2}{|\tilde{\gamma}| \tau_p}\sin\left({\frac{|z|}{\xi_p}}\right) + P_{z_B}&  \tilde{\gamma} <0 ,
\end{cases}\\
\end{array}
\ee
where $\tilde{\gamma}$ = $\gamma + \rho_p\omega_p^2$ and $P_{z_B}$ is the bulk solution. For $\tilde{\gamma}<0$, this interaction produces a modulation of the polarization. While $P_0$ = 0 for the FE case, the polarization effect on the superconductor is non-zero due to polarization interactions at the interface integrated over the screening length.

For the case $\tilde{\gamma}>$ 0, it is found that the inclusion of the linear interaction term {\em creates an interface-induced polarization} at $z$ = 0, which has an explicit decay length $\xi_p$ = $\sqrt{g_p/|\tilde{\gamma}| }$. The effect of fluctuations (finite $\omega_p$) will drive the $\tilde{\gamma}$ to be positive, which will push a FE state into a PE state. Using the same Fourier transformation for $z>$0, we find $\psi_{z}$ = $\psi_{z_I}$ + $\psi_{z_B}$ at the interface to be
\be
\begin{array}{c}
\displaystyle \psi_{z_I} =  \int_{-\infty}^{\infty} \frac{-2\lambda P_0\psi_0}{\tilde{\alpha} + g_{\psi}k_{\psi}^2}e^{ik_{\psi}z} dk_{\psi} + \psi_{z_B} \\ \\
=
\begin{cases}
\frac{2\lambda P_0\psi_0}{|\tilde{\alpha}| \tau_{\psi}}e^{-\frac{|z|}{\xi_{\psi}}} &\tilde{\alpha} >0 \\
\frac{2\lambda P_0\psi_0}{|\tilde{\alpha}| \tau_{\psi}}\cos \left({\frac{|z|}{\xi_{\psi}}}\right) + \psi_{z_B} &  \tilde{\alpha} <0 ,
\end{cases}\\
\end{array}
\ee
where $\tilde{\alpha}$ = $\alpha + \rho_{\psi}\omega_{\psi}^2$ and $\psi_{z_B}$ is the bulk solution. Similar to the FE state, this produces a modulation of $\psi$ at the interface due to $\tilde{\alpha}<$ 0. The difference for sin and cos comes from the condition that $\psi^* i \bigtriangledown \psi + i \bigtriangledown \psi^* \psi$ = 0. This also leads to a similar decay length $\xi_{\psi}$ = $\sqrt{g_{\psi}/|\tilde{\alpha}| }$ for the interface effect. Solving for $z$ = 0, we find expressions for the source terms of our transformations to be
\be
P_0 = \frac{\lambda \psi_0^2}{\sqrt{|\tilde{\gamma}| g_p}}=\frac{\sqrt{|\tilde{\alpha}| g_{\psi}}}{2\lambda}.
\ee
and
\be
\psi_0^2 = \frac{\sqrt{|\tilde{\alpha}| g_{\psi} |\tilde{\gamma}| g_p}}{2 \lambda^2},
\ee
which demonstrates the dependence of the interaction on the symmetry breaking at the interface. It should be noted that fluctuations of the SC state with finite $\omega_{\psi}$ will push $\tilde{\alpha}$ to a positive value, which will ultimately destroy the SC state. Note: These modulations are idealistic. In real materials, the modulations will decay due to temperature and disorder effects and provide a shape similar to that shown in Fig. \ref{Pdecay}.

Figures \ref{Pdecay}(a) and (b) illustrate the $z$-dependence of the order parameters at FE/SC and PE/SC interfaces in comparison to the bulk material for $\omega_p$=$\omega_{\psi}$=0. In Fig. \ref{Pdecay}(a), the FE polarization produces an induction of the SC order parameter, which in turn modulates the polarization of the FE state at the interface. Figure \ref{Pdecay}(b) shows the SC state interacting with a PE state, where the presence of $\psi_{z}$ creates an interface-induced polarization in the PE at the interface. The modulation in the SC state of PE/SC interface is due to reverse screening from the induced polarization.

We can obtain a rough estimate for the interaction coupling constant $\lambda$ through a comparison to the electric potential energy
\be
\frac{q_p q_{\psi}}{4\pi \epsilon_0d}=\lambda P_0\psi_0^2,
\ee
where $q_p$ is the surface charge for the polarization, $q_{\psi}$ is the surface charge generated by the SC, $\epsilon_0$ is the permittivity of free space, and $d$ is the interaction distance ($\sim$10\AA). The surface charge of the superconductor can be written as $q_{\psi}  = 0.02e,$ where $\rho_{\psi}$ is the charge density (1x10$^{15}$e/cm$^2$), $A_{\psi}$ is the surface area of the interface ($\sim(4$\AA)$^2$), $\Delta_{SC}$ is the SC gap (10meV), and $E_f$ is the fermi energy (100meV). We assume $q_{\psi}$ = $q_p$ and estimate the electric potential energy of this system to be on the order of 1 meV. Since $\psi_0$ is proportional to the SC gap, it can be estimated to about 10 meV. Therefore, if $P_0$ is on the order of 1 mC/m$^2$, then $\lambda \approx10 \frac{\rm{m}^2}{\rm{meV C}}$. This provides an order of magnitude estimate based from standard SC and electric polarization parameters with respect to the surface charge of the superconductor. If the electric polarization dramatically changes the SC electron density at the interface, then $\lambda$ can vary by an order of magnitude.

The presence of the interface-induced polarization should be experimentally observable. This could be achieved in heterostructures of a superconductor (YBCO)\cite{wu:87} and a PE (SrTiO$_3$)\cite{mull:79}  or a  FE where one deliberately induces fluctuations of polarization(BaTiO$_3$).\cite{smit:08} The latter technique would provide a direct comparison between FE and PE phases.

{\it Conclusion}:  By investigating the electronic states between electric polarization and superconducting interfaces, we find that the explicit interfacial symmetry breaking along the perpendicular direction enables a coupling that is linear in electric polarization $\lambda P_{z_I}|\psi_{z_I}|^2$. By investigating the effective action, it is found that gradient effects at the interface produce surface-induced modulations of the order parameters even in case when there are bulk ordered states. 

For the specific case of the PE/SC interface,  we find the PE state creates an interface-induced polarization at the interface which decays into a bulk. It should be mentioned that the induction of a SC from a non-SC state may also be possible through an interaction with the FE/PE states depending on the mechanism of the interaction. Details of this induction will be presented elsewhere.


We would like to acknowledgement helpful discussions with Lev Boulaevskii, Quanxi Jia, Dzmirty Yarotski and Jian-Xin Zhu.
This work was supported , in part, by the Center for Integrated Nanotechnologies, a U.S. Department of Energy, Office of Basic Energy Sciences user facility and in part by the LDRD and, in part, by UCOP TR-027.  Los Alamos National Laboratory, an affirmative action equal opportunity employer, is operated by Los Alamos National Security, LLC, for the National Nuclear Security Administration of the U.S. Department of Energy under contract DE-AC52-06NA25396.


\begin{thebibliography}{har}
\bibitem{linearnote}
Given the electro-static potential interaction $\int \rho_z |\psi_{z^{\prime}}|^2 K(z,z^{\prime}) dz dz^{\prime}$, where $\rho_z$ is the charge density and is proportional to $c_1P_0$ and $K(z,z^{\prime})$ is the electro-static kernel. Here, $\int |\psi_{z^{\prime}}|^2 K(z,z^{\prime}) dz^{\prime}$ is proportional to $\kappa |\psi_0|^2$. Therefore, the linear interaction term at the interface is derived as $c_1 \kappa P_0  |\psi_0|^2$ or $\lambda P_0  |\psi_0|^2$.
\end{thebibliography}
\end{document}